\documentclass[aps,prd,reprint,groupedaddress,showpacs]{revtex4-1}
\usepackage[margin=1in]{geometry}
\usepackage{graphicx, subfig}
\begin{document}

%\preprint{}

\title{Implications of $\delta^{CP}_l\sim 270^\circ$ and $\theta_{23}\gtrsim 45^\circ$ for texture specific lepton mass matrices and $0\nu \beta \beta$ decay}

\author{Rohit Verma}
\email[]{rohitverma@live.com}
\affiliation{Chitkara University, Barotiwala, Himachal Pradesh, 174103, India}
%\thanks{}
\date{\today}

\begin{abstract}
We study the phenomenological consequences of recent results from atmospheric and accelerator neutrino experiments, favoring normal neutrino mass ordering $m_1 < m_2 < m_3$, a near maximal lepton Dirac CP phase $\delta_{l}\sim 270^\circ$ along with $\theta_{23}\gtrsim 45^\circ$, for possible realization of natural structure in the lepton mass matrices characterized by ${({M_{ij}})} \sim O(\sqrt {{m_i}{m_j}}) $ for $i,j=1,2,3$. It is observed that deviations from parallel texture structures for $M_{l}$ and $M_{\nu}$ are essential for realizing such structures. In particular, such hierarchical neutrino mass matrices are not supportive for a vanishing neutrino mass $m_{\nu 1}\rightarrow 0$ characterized by Det$M_{\nu}\ne 0$ and predict ${m_{\nu 1}} \simeq (0.1 - 8.0)$ $meV$ , ${m_{\nu 2}} \simeq (8.0 - 13.0)$ $meV$, ${m_{\nu 3}} \simeq (47.0 - 52.0)$ $meV$, $\Sigma  \simeq (56.0 - 71.0)$ $meV$ and $\left\langle {{m_{ee}}} \right\rangle  \simeq (0.01 - 10.0)$ $meV$, respectively, indicating that the task of observing a $0\nu \beta \beta$ decay may be rather challenging for near future experiments. 

\end{abstract}

\maketitle
\section{Introduction}
One of the intriguing phenomenon in particle physics is the origin of fermion masses which appear to span several orders of magnitudes starting with neutrinos to the top quark. The masses and flavor mixing schemes of quarks and leptons are significantly different with the quark sector exhibiting strong mass hierarchy, small mixing angles and relatively heavier mass spectrum whereas the neutrinos are extremely light while two of their mixing angles are still large. In the current scenario, there is also a lack of consensus on the nature of neutrinos i.e. Dirac or Majorana along with doubts on the possible ordering of neutrino masses viz. normal ie. $m_1 < m_2 < m_3$ (NO) or inverted i.e. $m_3 < m_1 < m_2$ (IO). This nevertheless makes the task of constructing the fermion mass matrices non-trivial especially in the context of quark-lepton complementarity.

The confirmation of Higgs Boson by the ATLAS and CMS   collaborations \cite{Aad:2015zhl} completes the Standard Model (SM) of particle physics. Within this model, the quark  mass terms  in the Lagrangian are expressible as 
\begin{equation}
- L_{mass}^{quarks} = {{\bar q}_{uL}}{M_u}{q_{uR}} + {{\bar q}_{dL}}{M_d}{q_{dR}} + h.c.
\end{equation}
where ${q_{uL(R)}}$ and ${q_{dL(R)}}$ are the left(right) handed quark fields and $M_q$ are the quark mass matrices with $u,d$ for the "up" type and "down" type quarks. The resulting weak charged current quark interactions are given by
\begin{equation}
- {L_{cc}^{quarks}} = \frac{g}{{\sqrt 2 }}{\overline {\left( {\begin{array}{*{20}{c}}
u&c&t
\end{array}} \right)} _L}{\gamma ^\mu }V_{CKM}{\left( {\begin{array}{*{20}{c}}
d\\
s\\
b
\end{array}} \right)_L}W_\mu ^ +  + h.c.
\end{equation}
where $V_{CKM} = U_L^{{\rm{u}}\dag }U_L^{\rm{d}}$ is the Cabibbo-Kobayashi-Maskawa (CKM) matrix \cite{Cabibbo:1963yz, Kobayashi:1973fv} or the quark mixing matrix measuring the non-trivial mismatch between the flavor and mass eigenstates of quarks e.g.
\begin{equation}
\begin{array}{l}
U_L^{{\rm{u}}\dag }{M_{\rm{u}}}M_{\rm{u}}^\dag U_R^{\rm{u}} = Diag(\begin{array}{*{20}{c}}
{m_{\rm{u}}^2},&{m_{\rm{c}}^2},&{m_{\rm{t}}^2}
\end{array}),\\
U_L^{{\rm{d}}\dag }{M_{\rm{d}}}M_{\rm{d}}^\dag U_R^{\rm{d}} = Diag(\begin{array}{*{20}{c}}
{m_{\rm{d}}^2},&{m_{\rm{s}}^2},&{m_{\rm{b}}^2}
\end{array}).
\end{array}
\end{equation} 
where $U^q$ are unitary matrices.

Interestingly, the quark masses as well as the elements of CKM matrix, observe a hierarchical pattern viz. ${m_1}\ll {m_2}\ll{m_3}$ and $({V_{{\rm{ub}}}},{V_{{\rm{td}}}}) < ({V_{{\rm{cb}}}},{V_{{\rm{ts}}}})< ({V_{{\rm{us}}}},{V_{{\rm{cd}}}})<({V_{{\rm{ud}}}},{V_{{\rm{cs}}}},{V_{{\rm{tb}}}})$. It is natural to expect this hierarchy to be embedded within the corresponding quark mass matrices namely (for q = u,d)
\begin{equation}
M_{11} < M_{12,21}\lesssim M_{13,31} < M_{22} < M_{23,32} < M_{33},
\end{equation}
with $M_{22} << M_{33}$. Recent investigations \cite{Verma:2015mgd} in this regard indicate that the current quark mixing data indeed permit the quark mass matrices to have such a natural and hierarchical structure provided ${({M_{ij}})} \sim O(\sqrt {{m_i}{m_j}})$ for $i,j=1,2,3, i\ne j$ and ${({M_{ii}})} \sim O({{m_i}})$. Such hierarchical mass matrices have been referred to as \textit{natural mass matrices} \cite{Peccei:1995fg}. In particular, naturalness provides a rationale framework to correlate the observed fermion mass ratios, the corresponding mass matrices and observed mixing angles. Specifically, for $({M_{13}})=({M_{31}})\ne0$, the observed strong hierarchy among the quark masses and CKM elements gets naturally translated onto the structure of the corresponding mass matrices. 

A concomitant of such naturalness in mass matrices is the absence of parallel texture structure for the "up" and "down" type quark mass matrices \cite{Verma:2015mgd}. A \textit{parallel} texture structure corresponds, for example, to the mass matrices $M_u$ and $M_d$ with texture zeros at identical positions in both the mass matrices. Hierarchical structures have fetched greater importance in literature as these predict certain very simple yet compelling relations among the CKM elements and the quark mass ratios \cite{Verma:2015mgd, Fritzsch:1979zq, Fritzsch200363, Xing:2003yj, Hall1993164, Ramond:1993kv, Barbieri199993, Roberts2001358, Kim:2004ki, Branco:2006wv}. 

However, the mass spectrum for leptons is quite distinguished from the quark sector, wherein the charged leptons masses are strongly hierarchical i.e. $m_e \ll m_\mu \ll m_\tau$ while at least two of the neutrinos are allowed to have the same order of mass. It should be interesting to investigate if naturalness can provide a unique explanation for the fermion mass matrices, corresponding observed mass spectra as well as the mixing angles both for the quark as well as the lepton sectors.

Since the neutrinos are massless within the SM, one has to explore beyond the realms of SM to comprehend the origin of neutrino masses and observed neutrino oscillations phenomenon. A simplistic way to achieve this is to extended the SM theory by assuming neutrinos as Dirac-like particles. In this case, the neutrinos acquire mass through the Higgs mechanism in the similar way as quarks and charged leptons do within the SM, through a Dirac mass term e.g.
\begin{equation}
\begin{array}{l}
 - L_{mass}^{l} = {{\bar l}_L}{M_l}{l_R} + h.c.,\\\\
 - 2L_{mass}^{Dirac} = {{\bar \nu }_L}{M_{\nu_D}}{\nu _R} + h.c.,
\end{array}
\end{equation}
where $M_l$ and ${M_{\nu_D}}$ represent the charged-lepton and Dirac neutrino mass matrix respectively. Indeed, the current experiments have not ruled out such a possibility. In this context, it is also observed that highly suppressed Yukawa couplings for Dirac neutrinos can naturally be achieved using models with extra spatial dimensions \cite{Dienes199925,PhysRevD.65.024032} or through radiative mechanisms \cite{PhysRevD.18.1621,PhysRevLett.58.1600,PhysRevD.59.113008,1475-7516-2006-12-010,PhysRevD.77.105031,Gu201338}. However, such a possibility is perceived to be highly unlikely due to several orders of magnitude difference among ${m_\alpha }$ ($\alpha$=e,$\mu$,$\tau$) and ${m_{\nu i}}$ ($i$=1,2,3).

A rather convincing and natural explanation of neutrino masses can be obtained if neutrinos are assumed to be Majorana particles. This usually involves adding the lepton number (and flavor) violating Majorana mass terms for neutrinos in the Lagrangian e.g. 
\begin{equation}
 - 2L_{mass}^{Majorana} = {{\bar \nu }_L}{M_{\nu_L}}\nu _R^c + \bar \nu _L^c{M_{\nu_R}}{\nu _R} 
\end{equation}
where ${M_{\nu_L}}$ and ${M_{\nu_R}}$ correspond respectively to the left and right handed Majorana neutrino mass matrices, and the latter usually has an extremely high mass scale. This facilitates in generating the light neutrino masses through the Type-I or Type-II seesaw mechanisms viz. 
\begin{equation}
{M_\nu } =  - M_{\nu_D}^TM_{\nu_R}^{ - 1}{M_{\nu_D}}
\end{equation}
and 
\begin{equation}
{M_\nu } = {M_{\nu_L}} - M_{\nu_D}^TM_{\nu R}^{ - 1}{M_{\nu_D}},
\end{equation}
where $M_\nu$ is usually a complex symmetric matrix e.g.
\begin{equation}
{M_\nu } = \left( {\begin{array}{*{20}{c}}
{{e_\nu }}&{{a_\nu }}&{{f_\nu }}\\
{{a_\nu }}&{{d_\nu }}&{{b_\nu }}\\
{{f_\nu }}&{{b_\nu }}&{{c_\nu }}
\end{array}} \right).
\end{equation}

This allows writing the corresponding charged weak current term for leptons as
\begin{equation}
- {L_{cc}^{leptons}} = \frac{g}{{\sqrt 2 }}{\overline {\left( {\begin{array}{*{20}{c}}
{\nu_{e}}&{\nu_{\mu}}&{\nu_{\tau}}
\end{array}} \right)} _L}{\gamma ^\mu }V{\left( {\begin{array}{*{20}{c}}
e\\
{\mu}\\
{\tau}
\end{array}} \right)_L}W_\mu ^ +  + h.c.
\end{equation}
where $V = V_{PMNS}=U_{lL}^\dag {U_{\nu L}}$ is the Pontecorvo-Maki-Nakagawa-Sakata(PMNS) mixing matrix \cite{Maki01111962} or the neutrino mixing matrix and emerges through the diagonalization of the matrices $M_l$ and ${M_{\nu}}$, e.g.
\begin{equation}
\begin{array}{c}
U_{lL}^\dag{M_{l}}M_{l}^\dag U_{lR} = Diag(\begin{array}{*{20}{c}}
{m_{\rm{e}}^2},&{m_{\mu}^2},&{m_{\tau}^2}
\end{array}),\\\\
{U_{\nu L}}\dag {M_{\nu}}M_{\nu}^\dag {U_{\nu R}} = Diag(\begin{array}{*{20}{c}}
{m_{\nu_1}^2},&{m_{\nu_2}^2},&{m_{\nu_3}^2}
\end{array}).
\end{array}
\end{equation}
This mixing matrix relates the neutrino flavor states with the neutrino mass eigenstates through 
\begin{equation}
{\nu _{\alpha L}} = \sum\limits_{i = 1,2,3} {{V_{\alpha i}}{\nu _{iL}}}.
\end{equation}
In the standard parametrization \cite{Agashe:2014kda}, the PMNS matrix is expressed as $V = U \cdot P_o$, where $P_o \equiv Diag\{ {e^{i\rho }},{e^{i\sigma }},1\} $ with $\rho$, $\sigma$ being two Majorana CP violating phases and $U$ can be parametrized in terms of three mixing angles $\theta_{12}$,$\theta_{13}$,$\theta_{23}$ and one Dirac CP violating phase $\delta_{l}$ namely, 
\begin{widetext}
\begin{equation}
U = \left( {\begin{array}{*{20}{c}}
{{c_{12}}{c_{13}}}&{{s_{12}}{c_{13}}}&{{s_{13}}{e^{ - i{\delta _l}}}}\\
{ - {s_{12}}{c_{23}} - {c_{12}}{s_{13}}{s_{23}}{e^{i{\delta _l}}}}&{{c_{12}}{c_{23}} - {s_{12}}{s_{13}}{s_{23}}{e^{i{\delta _l}}}}&{{s_{23}}{c_{13}}}\\
{{s_{12}}{s_{23}} - {c_{12}}{c_{23}}{s_{13}}{e^{i{\delta _l}}}}&{ - {c_{12}}{s_{23}} - {s_{12}}{s_{13}}{c_{23}}{e^{i{\delta _l}}}}&{{c_{23}}{c_{13}}}
\end{array}} \right)
\end{equation}
\end{widetext}
with ${s_{ij}} = {\mathop{Sin}\nolimits} {\theta _{ij}}$ and ${c_{ij}} = {\mathop{ Cos}\nolimits} {\theta _{ij}}$ for $ij=12,13,23$. The neutrino oscillation experiments provide constraints on the three mixing angles $\theta_{12}$,$\theta_{13}$,$\theta_{23}$ along with the two mass square differences viz. $\delta {m^2} = m_2^2 - m_1^2$ and $\Delta {m^2} = \eta [m_3^2 - \frac{{(m_1^2 + m_2^2)}}{2}]$ with $\eta$=+1 for NO and $\eta$= -1 for IO cases.

In the current scenario, the global picture of neutrino oscillation parameters for NO at 3$\sigma$ suggests \cite{PhysRevD.86.013012}
\begin{equation}\label{data}
\begin{array}{c}
\delta {m^2} = (6.99 - 8.18) \times {10^{ - 5}}e{V^2},\\
\Delta {m^2} = (2.23 - 2.61) \times {10^{ - 3}}e{V^2},\\
s_{12}^2 = 0.259 - 0.359,\\
s_{13}^2 = 0.0176 - 0.0295,\\
s_{23}^2 = 0.374 - 0.626,\\
{\delta _l} (1\sigma) = {201^\circ } - {239^\circ }.
\end{array}
\end{equation}
Moreover, the above data does not seem to forbid $m_{\nu 1}=0$ for NO or $m_{\nu 3}=0$ for IO cases, the signatures for which are obtained through Det$M_{\nu}=0$.
The Planck collaboration measurements of the cosmic microwave background \cite{Ade:2015xua} provide further insight on the sum of absolute neutrino masses, e.g.
\begin{equation}
\Sigma  = {m_{\nu 1}} + {m_{\nu 2}} + {m_{\nu 3}} < 0.23{\rm{ }} eV.
\end{equation} 
More recent results from long-baseline accelerator neutrino experiments T2K \cite{Abe:2015awa} and NO$\nu$A \cite{Zhou:2015qua} are indicative of a near maximal Dirac CP phase i.e.
\begin{equation}
\begin{array}{c}
\delta_{l}\sim 270^\circ,\\\\
\theta_{23}\gtrsim 45^\circ
\end{array}
\end{equation}
along with preference for the normal ordering (NO) of neutrino masses. These results are also supported by the preliminary results from the atmospheric neutrino experiment at Super-Kamiokande \cite{Zhou:2015qua}. In addition, a statistical analysis of the cosmological data \cite{Dell'Oro:2015tia} also indicates preference for NO providing maximum likelihood for Majorana effective mass i.e.
\begin{equation}
\left\langle {{m_{ee}}} \right\rangle  < 16meV
\end{equation}
in neutrinoless double beta decay at 1$\sigma$ where 
\begin{equation}
\left\langle {{m_{ee}}} \right\rangle = \left| {{e^{i\rho }}\left| {U_{e1}^2} \right|{m_{\nu 1}} + {e^{i\sigma }}\left| {U_{e2}^2} \right|{m_{\nu 2}} + \left| {U_{e3}^2} \right|{m_{\nu 3}}} \right|.
\end{equation}

As the mixing angles are related to the corresponding mass matrices, it therefore becomes desirable to study the implications of a combination of NO, $\delta_{CP}\sim 270^\circ$ along with $\theta_{23}\gtrsim 45^\circ$ for lepton mass matrices assuming quarks and lepton mass matrices have similar origins and investigate the conditions affecting the possibility of obtaining natural lepton mass matrices, synchronous with the quark sector. Nevertheless, from a top-down prospective, it should be more economical to have a common framework explaining the fermion masses and mixing for the quark and lepton sectors. 
\section{Lepton mass matrices}
Phenomenologically, the problem of constructing the fermion mass matrices has always been a difficult task within the framework of Standard Model (SM) and its possible extensions, wherein the flavor structure of these matrices
is usually not constrained by the gauge symmetry. As a result, the matrices ${{{M}}_{l}}$ and ${{{M}}_{\nu}}$ remain arbitrary $3 \times 3$ complex matrices thereby involving several free parameters as compared to the number of physical observables, namely six lepton masses, three mixing angles and one Dirac-like CP phase $\delta_l$ along with two Majorana phases $\rho$ and $\sigma$. 

In this regard, the "texture zero" ansatz initiated by Weinberg \cite{TNYA:TNYA2958} and Fritzsch \cite{Fritzsch1977436,Fritzsch:1977vd} has been quite successful in explaining the fermion masses and mixing patterns \cite{Chiu:2000gw, Xing200330,Fukugita2003273,Fritzsch2009220,Gupta:2009ur, Verma:2009gf,Mahajan:2009wd,Ahuja:2009jj,Verma:2010jy,Fritzsch:2011qv,Fukugita2012294,Liu:2012axa,Fritzsch20131457,Verma:2013cza,Fakay:2013gf,Verma:2013qta,Wang:2014dka,Verma:2014lpa,Verma:2014woa,Fritzsch:2015foa}. However, one requires to handle all possible texture structures on a case to case basis. In this context, a common framework allowing for the study of such possibilities is more desirable. This is addressed in the following section. 
\section{Constructing the PMNS matrix}
In order to reconstruct the PMNS matrix, one requires to obtain the diagonalizing transformations for the corresponding mass matrices. To start with, for $q=l,\nu$, we consider the following texture one zero mass matrices as
\begin{equation}
\begin{array}{c}
M_q = \left( {\begin{array}{*{20}{c}}
e_q {e^{i{\psi_q}}}&{{a_q}{e^{i{\alpha_q}}}}&0\\
{{a_q}{e^{  i{\alpha_q}}}}&d_q{e^{i{\omega_q}}}&{{b_q}{e^{i{\beta_q}}}}\\
0&{{b_q}{e^{i{\beta_q}}}}&c_q{e^{i{\gamma_q}}}
\end{array}} \right),\\\
M'_q = \left( {\begin{array}{*{20}{c}}
0&{{a'_q}{e^{i{\alpha_q}}}}&{{f'_q}{e^{i{\Delta_q}}}}\\
{{a'_q}{e^{i{\alpha_q}}}}&d'_q{e^{i{\omega_q}}}&{{b'_q}{e^{i{\beta_q}}}}\\
{{f'_q}{e^{i{\Delta_q}}}}&{{b'_q}{e^{i{\beta_q}}}}&c'_q{e^{i{\gamma_q}}}
\end{array}} \right).
\end{array}
\end{equation}
referred to as Type-I and Type-II texture structures respectively, in the following text.

One may also consider these matrices to be Hermitian for Dirac neutrinos. Using standard procedures, it is not possible to obtain the exact diagonalizing transformations for the latter case. In order to avoid a large number of free parameters in these matrices, we assume that the phases are factorizable in these, requiring 
\begin{equation}
\psi_q = 2\alpha_q, \omega_q = 0, \Delta_q=\alpha_q + \beta_q, \gamma_q = 2\beta_q
\end{equation}
for symmetric ${M_q}$ and ${M'_q}$ and 
\begin{equation}
\psi_q = 0, \omega_q = 0, \Delta_q=\alpha_q + \beta_q, \gamma_q = 0
\end{equation}
for Hermitian ${M_q}$ and ${M'_q}$. 

The diagonalization of ${M_q}$ above is realized using 
\begin{equation}
M_{q}^{Diag} = O_{q}^{T}{{\tilde M}_q}{O_q} = Diag\left( {m_1, - m_2, m_3} \right), 
\end{equation}
with $1,2,3=e,\mu,\tau$ for $q=l$ and $1,2,3={\nu 1},{\nu 2},{\nu 3}$ for $q=\nu$. Here  
\begin{equation}
P_q =  Diag\left( {e^{ - i\alpha_q},1,e^{ -i\kappa\beta_q}} \right)
\end{equation}
and
\begin{equation}
{{\tilde M}_q} = {P_q}{M_q}{Q_q} = \left( {\begin{array}{*{20}{c}}
{{e_q}}&{{a_q}}&0\\
{{a_q}}&{{d_q}}&{{b_q}}\\
0&{{b_q}}&{{c_q}}
\end{array}} \right),
\end{equation}
and $Q=P$ (symmetric case) and $Q=P^{\dag}$ (Hermitian case).
Considering $e_q$ and $d_q$ as free parameters, one can write \cite{Verma:2013qta} 
\begin{widetext}
\begin{equation}
{O_q}{\rm{ }} = {\rm{ }}\left( {\begin{array}{*{20}{c}}
{\sqrt {\frac{{({e_q} + {\rm{ }}{m_2})({m_3} - {e_q})({c_q} - {m_1})}}{{({c_q} - {e_q})({m_3} - {m_1})({m_2} + {\rm{ }}{m_1})}}} }&{\sqrt {\frac{{({m_1} - {e_q})({m_3} - {e_q})({c_q} + {\rm{ }}{m_2})}}{{({c_q} - {e_q})({m_3} + {\rm{ }}{m_2})({m_2} + {\rm{ }}{m_1})}}} }&{\sqrt {\frac{{({m_1} - {e_q})({e_q} + {\rm{ }}{m_2})({m_3} - {c_q})}}{{({c_q} - {e_q})({m_3} + {\rm{ }}{m_2})({m_3} - {\rm{ }}{m_1})}}} }\\
{\sqrt {\frac{{({m_1} - {e_q})({c_q} - {m_1})}}{{({m_3} - {m_1})({m_2} + {\rm{ }}{m_1})}}} }&{ - \sqrt {\frac{{({e_q} + {\rm{ }}{m_2})({c_q} + {\rm{ }}{m_2})}}{{({m_3} + {\rm{ }}{m_2})({m_2} + {\rm{ }}{m_1})}}} }&{\sqrt {\frac{{({m_3} - {e_q})({m_3} - {c_q})}}{{({m_3} + {\rm{ }}{m_2})({m_3} - {m_1})}}} }\\
{ - \sqrt {\frac{{({m_1} - {e_q})({m_3} - {c_q})({c_q} + {\rm{ }}{m_2})}}{{({c_q} - {e_q})({m_3} - {m_1})({m_2} + {\rm{ }}{m_1})}}} }&{\sqrt {\frac{{({e_q} + {\rm{ }}{m_2})({c_q} - {m_1})({m_3} - {c_q})}}{{({c_q} - {e_q})({m_3} + {\rm{ }}{m_2})({m_2} + {\rm{ }}{m_1})}}} }&{\sqrt {\frac{{({m_3} - {e_q})({c_q} - {m_1})({c_q} + {\rm{ }}{m_2})}}{{({c_q} - {e_q})({m_3} + {\rm{ }}{m_2})({m_3} - {m_1})}}} }
\end{array}} \right)
\end{equation}
\end{widetext}
such that
\begin{equation}
\begin{array}{c}
{c_q} = {\rm{ }}{m_1} - {m_2} + {\rm{ }}{m_3} - {d_q} - {e_q},{\rm{ }}\\
{a_q} = \sqrt {\frac{{\left( {{m_1} - {e_q}} \right)\left( {{m_2} + {\rm{ }}{e_q}} \right)\left( {{m_3} - {e_q}} \right)}}{{\left( {{c_q} - {e_q}} \right)}}} ,{\rm{ }}\\
{b_q} = \sqrt {\frac{{\left( {{c_q} - {m_1}} \right)\left( {{m_3} - {\rm{ }}{c_q}} \right)\left( {{c_q} + {\rm{ }}{m_2}} \right)}}{{\left( {{c_q} - {e_q}} \right)}}},\\
{m_1}{\rm{ }} > {\rm{ }}{e_q}{\rm{ }} > {\rm{ }} - {m_2},{\rm{ }}\\
({m_3} - {m_2} - {e_q}){\rm{ }} > {\rm{ }}{d_q}{\rm{ }} > {\rm{ }}({m_1} - {m_2} - {e_q}).
\end{array}
\end{equation}
The above constraints on the parameters $e_q$ and $d_q$ nevertheless allow hierarchical  mass matrices i.e. $e_q < a_q < d_q < b_q < c_q$. Texture rotation from the (13),(31) positions in  $M_q$ to (11) position in $M'_q$ is realized by rotating the (11) element in  $M_q$ to the (13),(31) position in $M'_q$ through a unitary transformation ${{{R}}_{\rm{q}}}$ on ${{{M}}_{{q}}}$ using 
\begin{equation}
M_q \to M'_q = R_{q}^T M_q R_q,
\end{equation}
for symmetric mass matrices and
\begin{equation}
M_q \to M'_q = R_{q}^\dag M_q R_q,
\end{equation}
for Hermitan case, where ${{{R}}_{\rm{q}}}$ is a complex rotation matrix in the 1-3 generation plane e.g.
\begin{widetext}
\begin{equation}
R_q = \left( {\begin{array}{*{20}{c}}
{{\mathop{\rm Cos}\nolimits} {\eta_{13_q}}}&0&{ - {e^{-i{(\alpha _q - \kappa\beta_q)}}}{\mathop{\rm Sin}\nolimits} {\eta_{13_q}}}\\
0&1&0\\
{e^{i{(\alpha _q - \kappa\beta_q)}}}{\mathop{\rm Sin}\nolimits} {\eta _{13_q}}&0&{{\mathop{\rm Cos}\nolimits} {\eta _{13_q}}}
\end{array}} \right).
\end{equation}
\end{widetext}
where $\kappa=+1$ for symmetric matrices and  $\kappa=-1$ for Hermitian matrices.  

The condition of a texture zero rotation from the (13,31) positions in $M_q$ to the (11) position in $M'_q$ requires 
\begin{equation}
0 = {e_q}{{\mathop{\rm Cos}\nolimits} ^2}{\eta _{{{13}_q}}} + {c_q}{{\mathop{\rm Sin}\nolimits} ^2}{\eta _{{{13}_q}}},
\end{equation}
which can be translated to
\begin{equation}
{{\mathop{\rm Tan}\nolimits} ^2}{\eta _{13}}_q = {{ - {e_q}} \mathord{\left/
 {\vphantom {{ - {e_q}} {{c_q}}}} \right.
 \kern-\nulldelimiterspace} {{c_q}}} \Rightarrow 
{\mathop{\rm Tan}\nolimits} {\eta _{13}}_q = {\tau _q}\sqrt {{{ - {e_q}} \mathord{\left/
 {\vphantom {{ - {e_q}} {{c_q}}}} \right.
 \kern-\nulldelimiterspace} {{c_q}}}} 
\end{equation}
where ${\tau _q} =  \pm 1 $ and $e_q$ is always negative. Note that the rotation angle ${\eta _{13}}_q$ is not a free parameter and is completely fixed through $e_q$ and $c_q$ due to repositioning of texture zeros as a result of the rotation $R_q$. One can now relate the matrix elements in $M'_	q$ with the corresponding elements in $M_q$, e.g.
\begin{equation}\label{32}
\begin{array}{c}
{{a'}_q} = \vert{a_q}{\mathop{\rm Cos}\nolimits} {\eta _{{{13}_q}}} + {\tau _q}{b_q}{\mathop{\rm Sin}\nolimits} {\eta _{{{13}_q}}}\vert,\\
{{b'}_q} = \vert{b_q}{\mathop{\rm Cos}\nolimits} {\eta _{{{13}_q}}} - {\tau _q}{a_q}{\mathop{\rm Sin}\nolimits} {\eta _{{{13}_q}}}\vert,\\
{{c'}_q} = {c_q}{{\mathop{\rm Cos}\nolimits} ^2}{\eta _{{{13}_q}}} + {e_q}{{\mathop{\rm Sin}\nolimits} ^2}{\eta _{{{13}_q}}},\\
{{d'}_q} = {d_q},
{{f'}_q} = \vert\sqrt { - {e_q}{c_q}}\vert .
\end{array}
\end{equation}

The texture rotation in 1-3 generation plane allows $d'_q=d_q$. Note that ${f'_q}\propto \sqrt { - e_q}$, while the other off-diagonal elements essentially get re-scaled due to texture rotation. Furthermore, for $e_q\sim -m_1$, one expects $f'_q\sim O(\sqrt{m_1 m_3})$ allowing hierarchical structures in the Type-II possibility namely $ a'_l <f'_l < d'_l < b'_l < c'_l$ along with $ a'_\nu \sim f'_\nu  \sim d'_\nu  \lesssim b'_\nu  \lesssim c'_\nu$ since $O(\sqrt{m_{\nu 1}m_{\nu 2}}) \sim O(\sqrt{m_{\nu 1}m_{\nu 3}}) \sim O(m_{\nu 2})$ are allowed by oscillation data. Henceforth, it is trivial to obtain the orthogonal transformation ${O'_{\rm{q}}}$ for ${M'_{\rm{q}}}$ (symmetric case) as
\begin{equation}
O'_q = P_{q} R_{q}^T P_q^{\dag} O_q = \tilde R_{q}^T O_q
\end{equation}
and (Hermitian case) as
\begin{equation}
O'_q = P_{q}^\dag R_{q}^\dag P_q O_q = \tilde R_{q}^T O_q
\end{equation}
with
\begin{equation}
M_{q}^{\prime Diag}= O_{q}^{\prime T} {{\tilde M'}_q} O'_q = M_{q}^{Diag}.
\end{equation}
with ${{\tilde M'}_q} = {P_q}M'{Q_q}$. Note that in the absence of texture rotation,  $\tilde R_q =I$ (unit matrix) for $M_q$ while
\begin{equation}
\tilde R_q = \left( {\begin{array}{*{20}{c}}
{{\mathop{\rm Cos}\nolimits} {\eta_{13_q}}}&0&{ - {\mathop{\rm Sin}\nolimits} {\eta_{13_q}}}\\\\
0&1&0\\
{\mathop{\rm Sin}\nolimits} {\eta _{13_q}}&0&{{\mathop{\rm Cos}\nolimits} {\eta _{13_q}}}
\end{array}} \right)
\end{equation}
for $M'_q$ signifying the corresponding effect of such rotation on real diagonilizing transformation $O'_q$. The resulting mixing matrix for $M_q$ and/or $M'_q$ may be constructed as
\begin{equation}\label{37}
V=O_{l}^T {\tilde R}_l P_l {P}_{\nu}^{\dag }{\tilde R}_{\nu}^T {O_\nu}. 
\end{equation}
Also $P_l P_{\nu}^\dag =Diag(e^{-i \phi_1},1,e^{i \phi_2}) $, $\phi_1 = \alpha_l - \alpha_\nu$ and $\phi_2 = \beta_\nu - \beta_l$ (symmetrics case) or $\phi_2 = \beta_l - \beta_\nu$ (Hermitian case). Note that a change in sign for $a'_q$ and $f'_q$ can always be accommodated in the redefinition of the phases $\alpha_q$ and $\beta_q$ which only appear implicitly in the PMNS matrix through $\phi_1$ and $\phi_2$. Considering the six lepton masses, $\phi_1$, $\phi_2$, $d_q$ and $e_q$ as free parameters, one can reconstruct the unitary mixing matrix $V$ using the above procedure and confront it with the current oscillation data. In lieu of this, we restrict our investigation to only texture four zero mass matices involving ten free parameters. Furthermore, the condition of naturalness forbids a texture zero at the (33) matrix elements.

Recent works \cite{Ludl:2014axa,Liao:2015hya,Ludl:2015lta,doi:10.1142/S0217732315300256} in this regard suggest that there exist several viable texture structures of lepton mass matrices. Most of these investigations work in the flavor basis with diagonal charged lepton mass matrix or enforce parallel texture structures for lepton mass matrices $M_l$ and $M_\nu$. In this letter, we investigate all possible structures for four zero lepton mass matrices, both symmetric and/or Hermitian, assuming factorizable phases (for simplicity) in these. The resulting structures are summarized in Tables 1 and 2 wherein we enlist all texture five and four zeros in agreement with current data at 3$\sigma$. The $X_l$ and $X_\nu$ in the tables represent the position of texture zeros in the corresponding mass matrices. It is observed that the constraints of  naturalness, near maximal $\delta_l$, $s_{23}^2 \gtrsim 0.50 $ and normal ordering for neutrino masses, taken together, greatly reduce the number of possible viable structures and only a few possibilities seem to survive the test. The possibility of a vanishing neutrino mass is also studied for these texture structures.  
\section{Fritzsch-like four zeros}
It has been observed \cite{Verma:2014woa,Verma:2013cza} that in the absence of $\delta_l \sim270^\circ$ constraint, the Fritzsch-like texture four zero mass matrices are physically equivalent to the generic lepton mass matrices. Interestingly, these matrices can be obtained from the above structures using the assumption of $e_q=0$ and $f'_q=0$. In particular, $R_q =\tilde R_q =I$, where $I$ is a unit matrix, for this case. The predictions from these matrices and their experimental tests can be found in previous works.  To start with, using Eqs.(\ref{data}), (\ref{37}) and allowing free variations to the parameters $m_{\nu{1}}$, ${d_l},{d_\nu },{\phi _1}$ and ${\phi _2}$, we first reconstruct the  viable structures for ${{\tilde M}_l}$ (in units of GeV) and ${{\tilde M}_\nu}$ (in units of eV) for $d_\nu \sim m_{\nu 2}$ using the available oscillation data and obtain the following best-fits:
\begin{widetext}
\begin{equation}\label{A}
\begin{array}{c}
{{\tilde M}_l} = \left( {\begin{array}{*{20}{c}}
0&{0.007 - 0.010}&0\\
{0.007 - 0.010}&{0 - 0.822}&{0.423 - 0.924}\\
0&{0.423 - 0.924}&{0.822 - 1.644}
\end{array}} \right)GeV,\\
{{\tilde M}_\nu } = \left( {\begin{array}{*{20}{c}}
0&{0.0066 - 0.0104}&0\\
{0.0066 - 0.0104}&{0.0076 - 0.0115}&{0.0223 - 0.0260}\\
0&{0.0223 - 0.0260}&{0.0302 - 0.0383}
\end{array}} \right)eV,
\end{array}
\end{equation}
\end{widetext}
along with ${\phi _1} = 0^\circ - 50^\circ,267^\circ - 360^\circ$ and ${\phi _2} = 180^\circ - 285^\circ$. The corresponding predictions for the absolute neutrino masses, $\Sigma$ and $\left\langle {{m_{ee}}} \right\rangle $ read ${m_{\nu 1}} = (2.96 - 6.70)$ $meV$, ${m_{\nu 2}} = (9.05 - 11.50)$ $meV$, ${m_{\nu 3}} = (47.7 - 51.9)$ $meV$, $\Sigma  = 60.2 - 69.6$ $meV$ and $\left\langle {{m_{ee}}} \right\rangle  = 0.008 - 9.00$ $meV$ respectively. In the context of agreement with $\delta_{l}\sim 270^\circ$ along with $\theta_{23}\gtrsim 45^\circ$,  it is observed that naturalness is allowed in $M_\nu$ independent of the $s_{23}$ octant. This is depicted in FIG. 1 where one observes that $d_\nu \lesssim m_{\nu 2}$ is still consistent with $s_{23}^2\gtrsim 0.5$. However, one finds that the near maximal constraint of $\delta_{l}\simeq 270^\circ$ requires large deviation of $M_l$ from a possible natural structure. In particular, we identify three vital sources for CP violation in these matrices namely the two non-trivial phases $\phi_1$, $\phi_2$ along with the free parameter $d_l$ as elaborated in FIG. 2. indicating $d_l>0.6$ GeV $\sim m_\tau /3 >> m_\mu$ is required to obtain $\delta_{l}\simeq 270^\circ$.  This also implies that Fritzsch-like texture five zero matrices  ($d_l=0$) should be ruled out by $\delta_{l}\simeq 270^\circ$. Our study reveals this conclusion to hold true for all possible texture five zero structures, all of which seem to be ruled out by a near maximal $\delta_l$, see Table 1. This calls upon investigating alternate texture structures, which on one hand account for near maximal $\delta_l$, and at the same time allow possible natural structures for $M_l$ and $M_\nu$ (i.e. $M_{jk}\sim O(m_j m_k)$).
\begin{figure}
\includegraphics[scale=1.0]{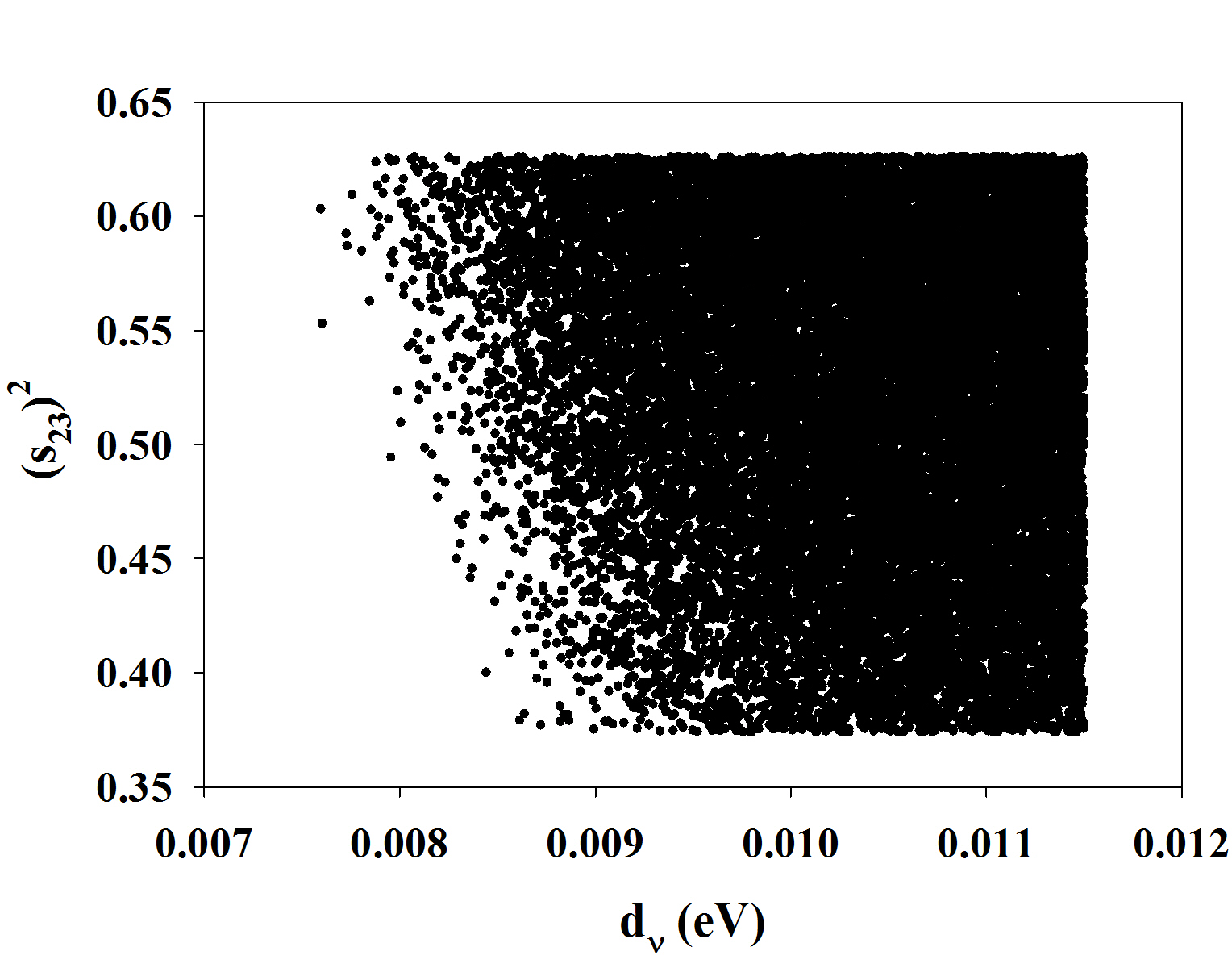}
\caption{$s_{23}^2$ vs. $d_\nu$ for Fritzsch-like four zeros.}
\end{figure}
\begin{figure}
\includegraphics[scale=1.0]{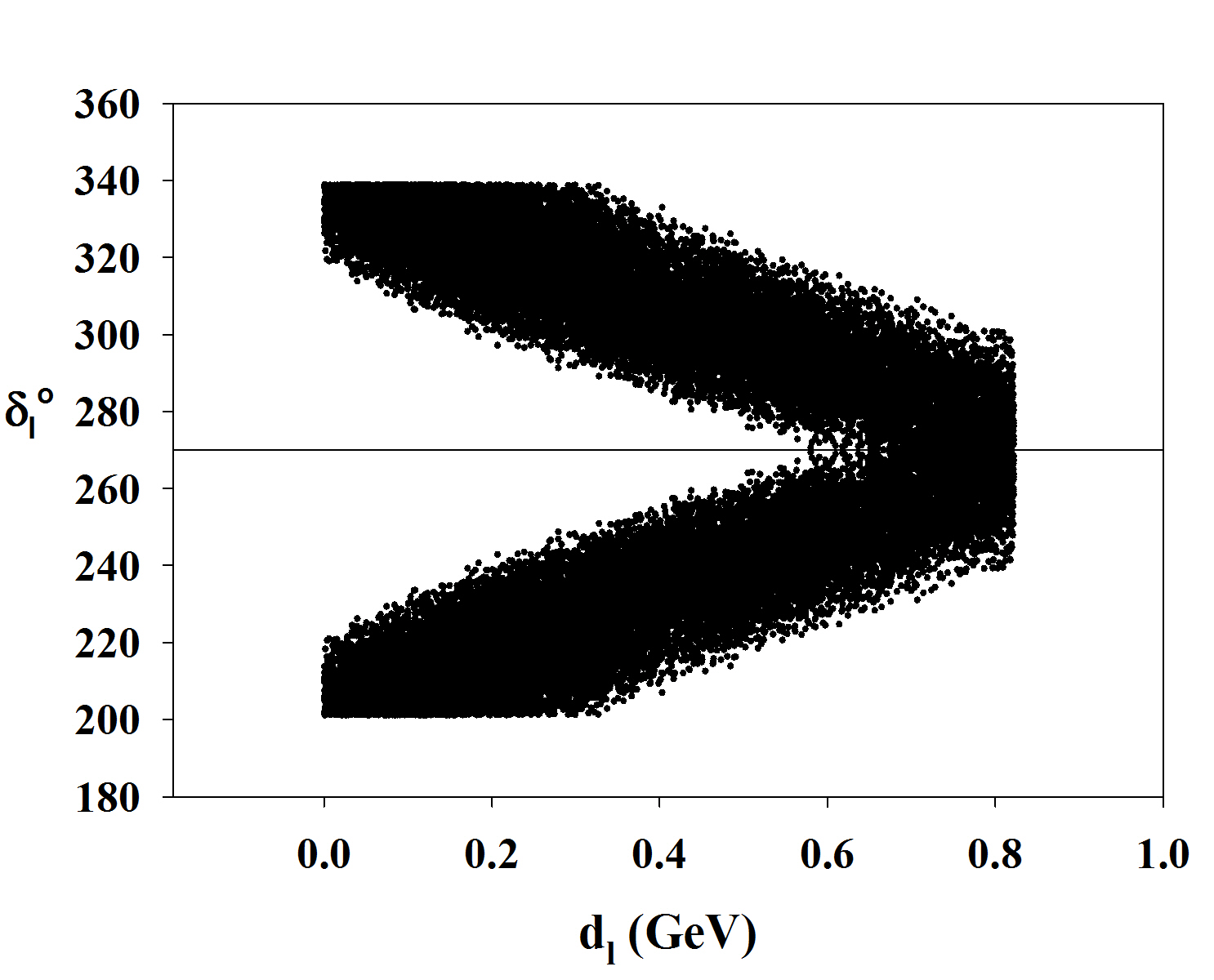}
\caption{$\delta_{l}$ vs. $d_l$ for Fritzsch-like four zeros.}
\end{figure}
\begin{figure}
\includegraphics[scale=1.0]{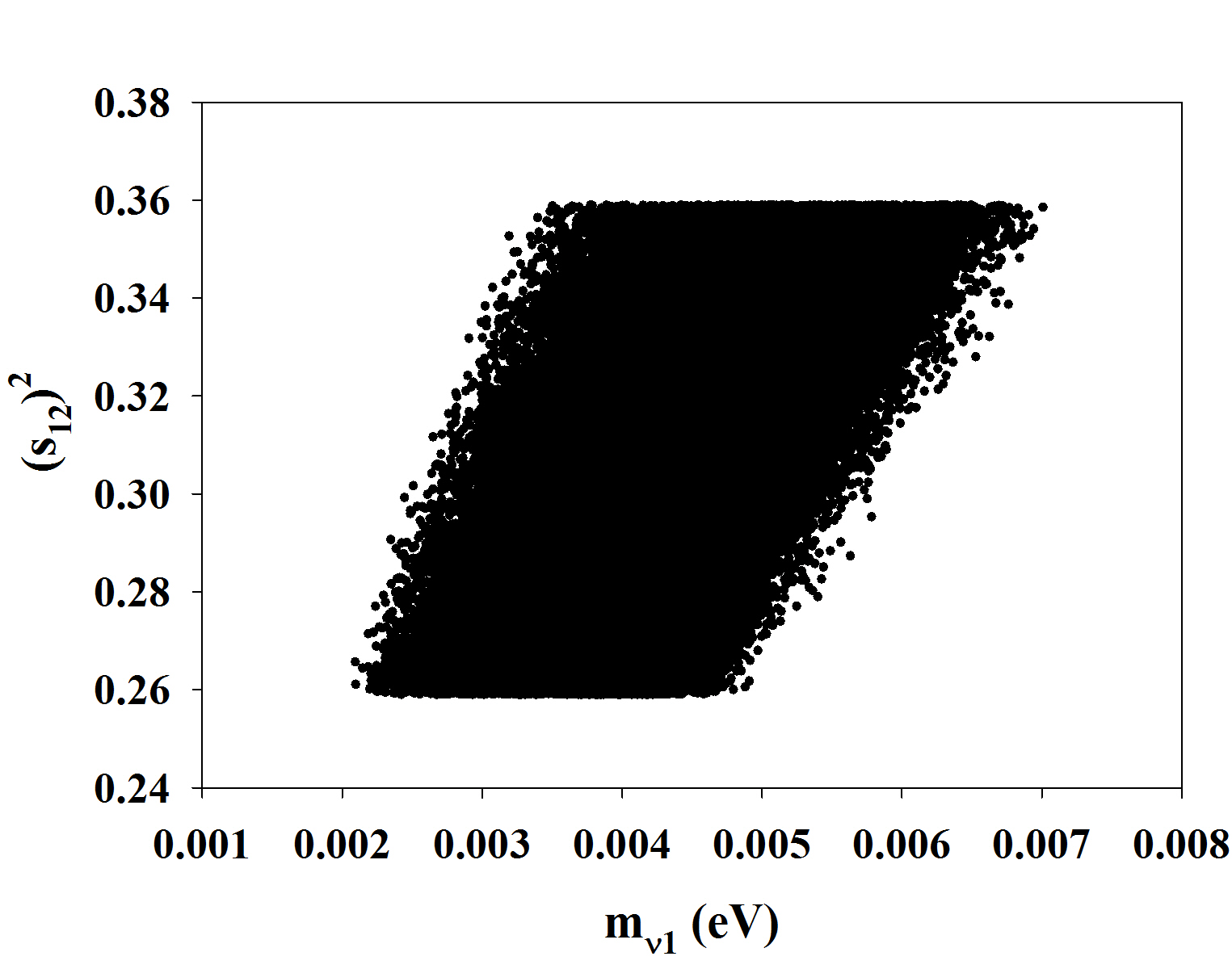}
\caption{$s_{12}^2$ vs. $m_{\nu 1}$ for Case-A.}
\end{figure}
\begin{figure}
\includegraphics[scale=1.0]{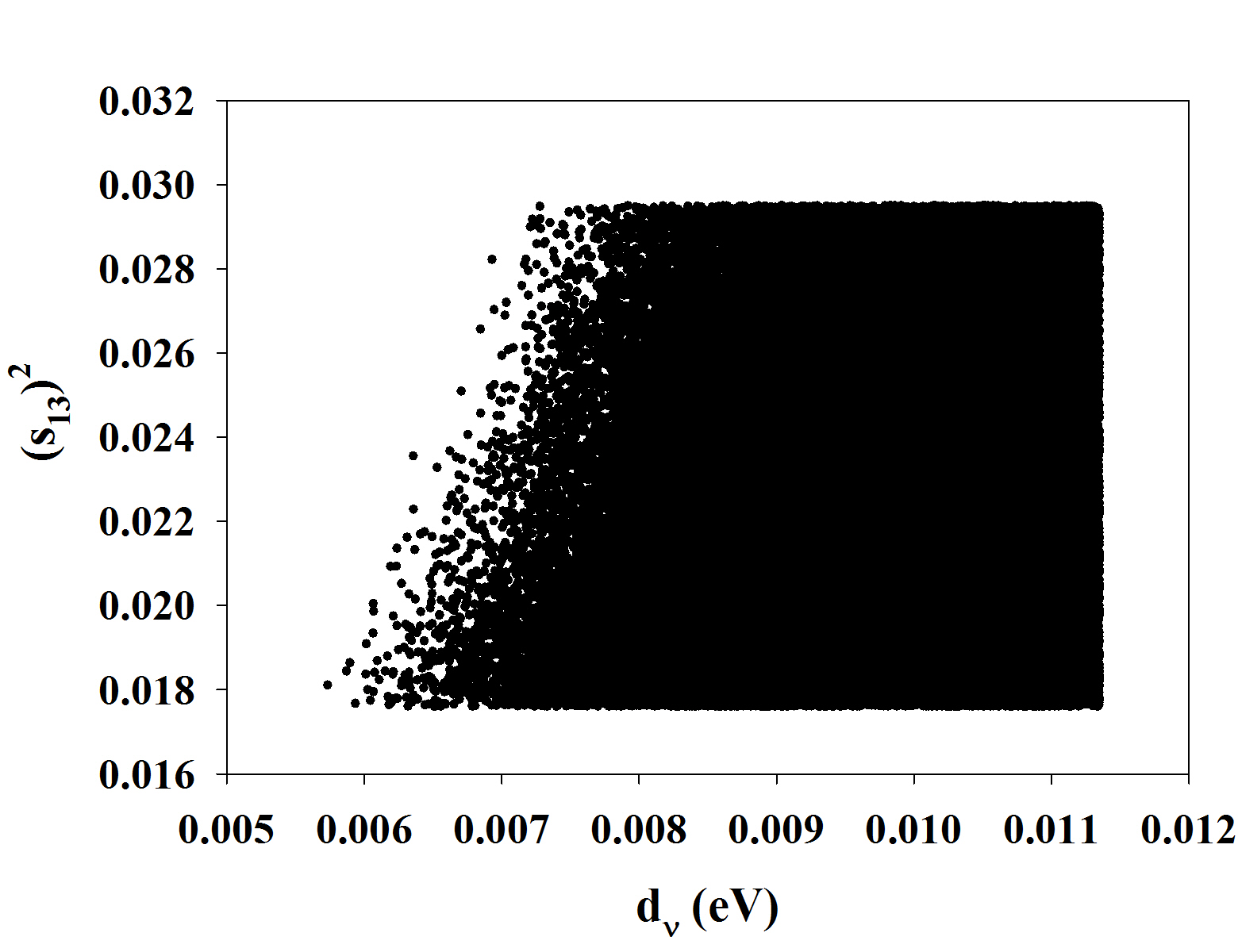}
\caption{$s_{13}^2$ vs. $d_{\nu}$ for Case-A.}
\end{figure} 
\begin{figure}
\includegraphics[scale=1.0]{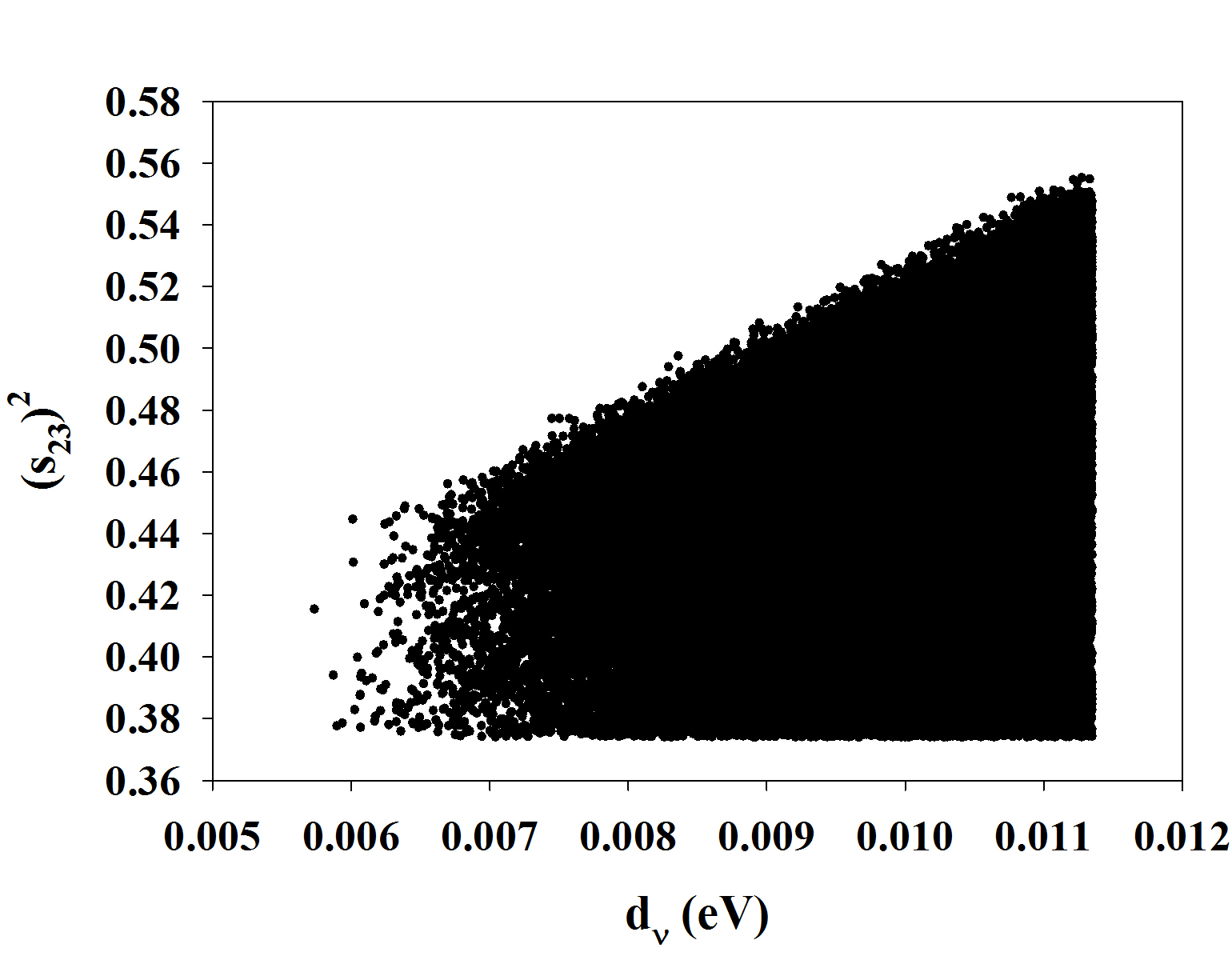}
\caption{$s_{23}^2$ vs. $d_{\nu}$ for Case-A.}
\end{figure}
\begin{figure}
\includegraphics[scale=1.0]{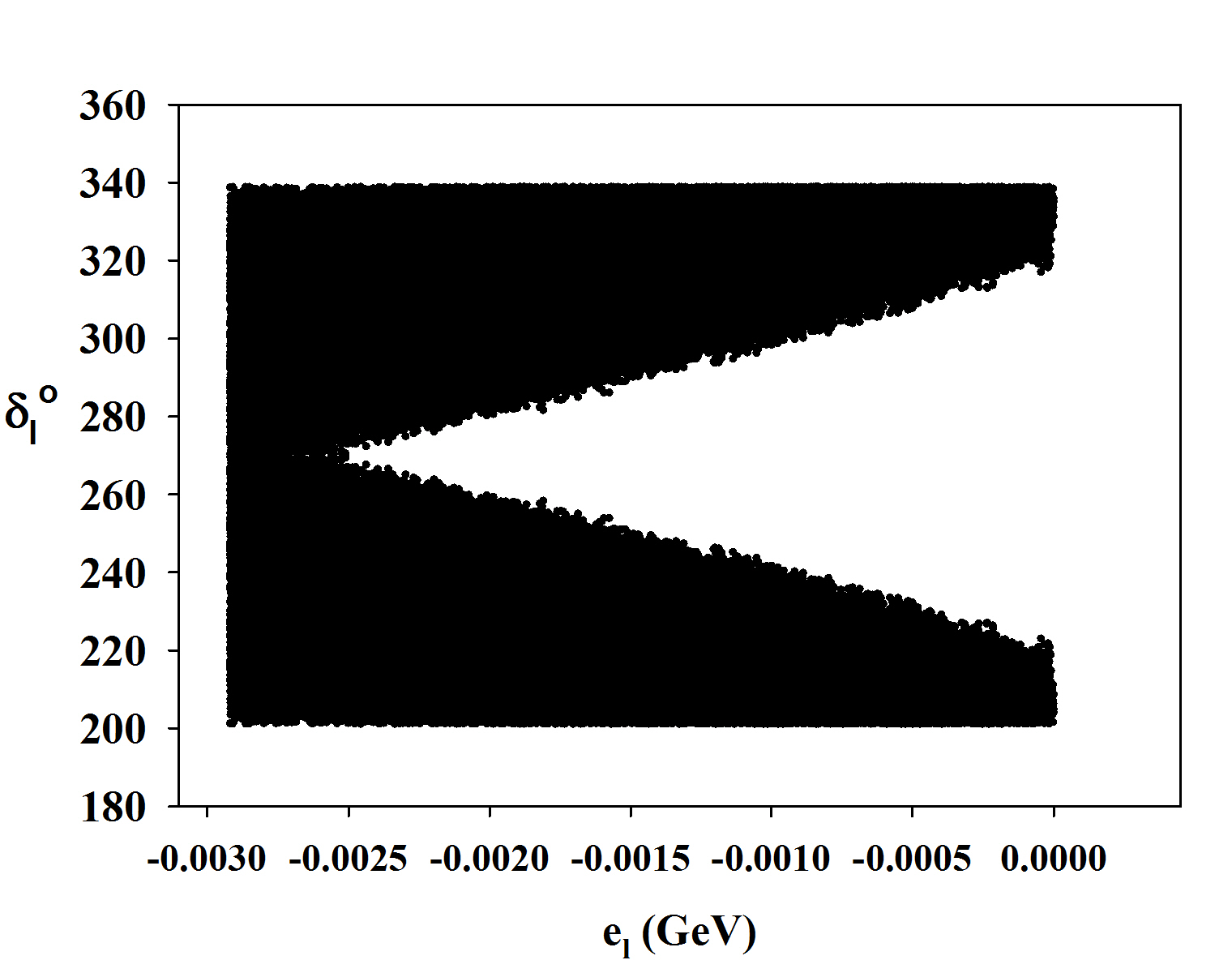}
\caption{$\delta_l$ vs. $e_l$ for Case-A.}
\end{figure}
\begin{figure}
\includegraphics[scale=1.0]{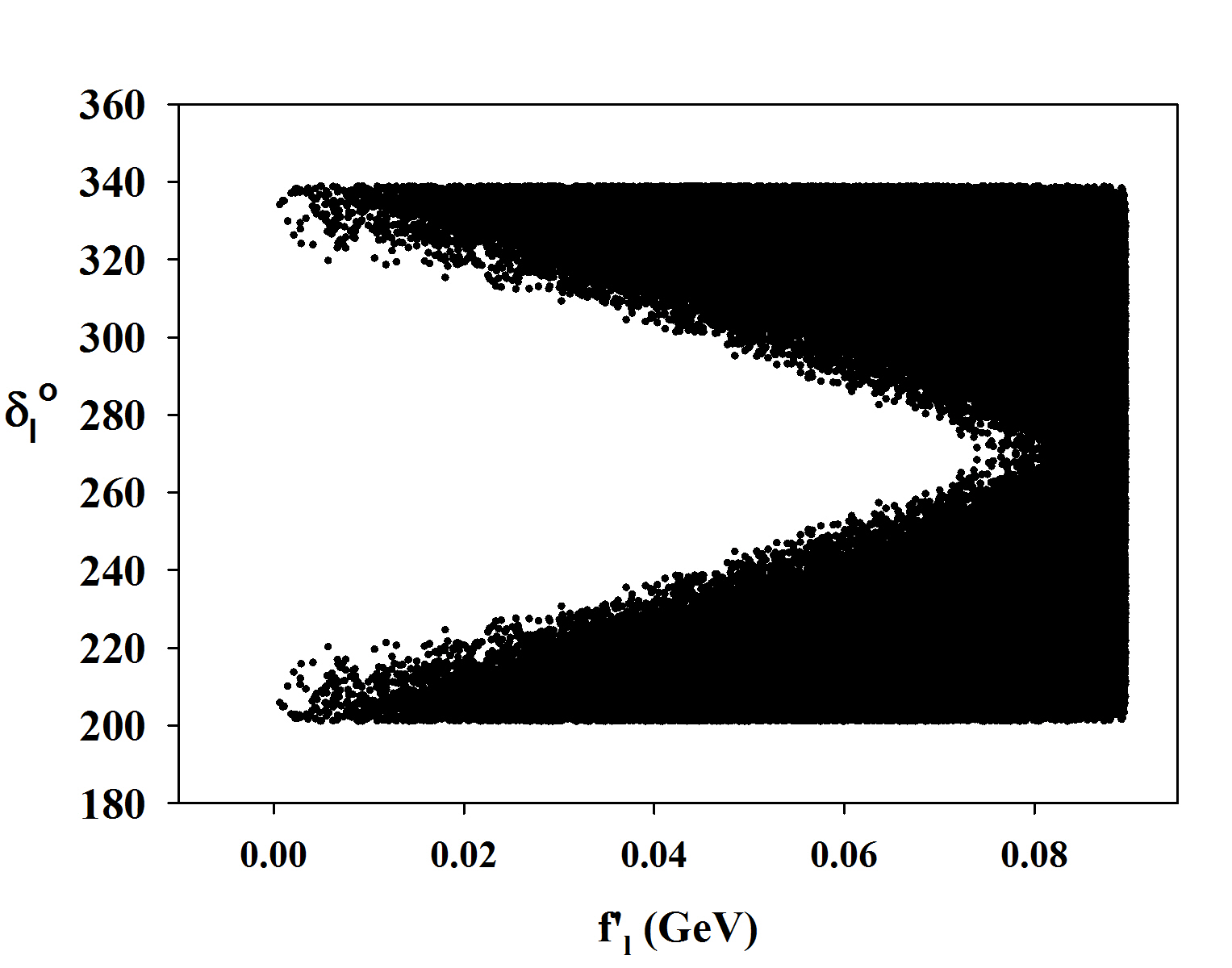}
\caption{$\delta_l$ vs. $f'_l$ for Case-B.}
\end{figure}
\section{Natural lepton mass matrices}
In this context, $e_q\ne0$ and/or $f'_q\ne0$ in the corresponding mass matrices provide greater possibility of realizing naturalness in corresponding mass matrices as compared to the Fritzsch-like structures wherein interactions between the first and third generation of leptons are suppressed due to texture zeros invoked at (11) and (13,31) matrix elements. At least, for the quark sector, non-vanishing (13,31) elements are observed to be crucial in effectuating the natural structures of corresponding mass matrices. A careful analysis of all possible texture four zero structures reveals that only four possibilities for natural structures are allowed by recent data, see Table 2. We categorize these as Type-I and Type-II, based on the texture structure of $M_\nu$.
\subsection{Type-I $M_\nu (11) = M_\nu (13,31)= 0$}
\subsubsection{Case-A $M_l (22) = M_l (13,31)= 0$}
The viable best-fit structures of the lepton mass matrices are summarized below
\begin{widetext}
\begin{equation}\label{B}
\begin{array}{c}
{{\tilde M}_l} = \left( {\begin{array}{*{20}{c}}
{-0.003-0}&{0.007 - 0.019}&0\\
{0.007 - 0.019}&0&{0.416 - 0.426}\\
0&{0.416 - 0.426}&{1.644-1.647}
\end{array}} \right)GeV,\\\\
{{\tilde M}_\nu } = \left( {\begin{array}{*{20}{c}}
0&{0.0053 - 0.0106}&0\\
{0.0053 - 0.0106}&{0.0057 - 0.0123}&{0.0221 - 0.0272}\\
0&{0.0221 - 0.0272}&{0.0285 - 0.0394}
\end{array}} \right)eV,
\end{array}
\end{equation}
\end{widetext}
with ${\phi _1} = 0^\circ - 340^\circ$, ${\phi _2} = 98^\circ - 265^\circ$ and ${m_{\nu 1}} = (1.99 - 7.01)$ $meV$, ${m_{\nu 2}} = (8.65 - 11.3)$ $meV$, ${m_{\nu 3}} = (47.7 - 51.9)$ $meV$, $\Sigma  = (58.7 - 70.0)$ $meV$ and $\left\langle {{m_{ee}}} \right\rangle  = (0.01 - 9.23)$ $meV$ respectively.  Like the Fritzsch-like texture four zeros,  $ s_{12}^2 \propto m_{\nu 1}$ \cite{Fritzsch:2015gxa,Fritzsch2009220,Verma:2014woa,Verma:2013cza} as depicted in FIG. 3. However, the other two mixing angles are fixed by the free parameter $d_\nu$ illustrated in FIGs. 4 and 5. The latter also indicates that natural structure for $M_\nu$ is allowed independent of the $s_{23}$ octant, with $d_\nu \lesssim m_{\nu 2}$ also accounting for $s_{23}^2>0.5$.  Finally, the parameter $e_l << m_\mu$ accounts  for near maximal $\delta_l$ as shown in the FIG. 6. In particular a small deviation of $\delta_l \longrightarrow 270^\circ \pm 30^\circ$ provides greater agreement of $e_l \sim 5 MeV$ with the notion of naturalness in the corresponding mass matrix.
\subsubsection{Case-B $M_l (11) = M_l (22)= 0$}
We obtain the following viable best-fit structures for these lepton mass matrices, namely 
\begin{widetext}
\begin{equation}\label{C}
\begin{array}{c}
{{\tilde {M}^\prime}_l} = \left( {\begin{array}{*{20}{c}}
0&{0.001 - 0.007}&0.0003-0.089\\
{0.001 - 0.007}&0&{0.413 - 0.423}\\
0.0003-0.089&{0.413 - 0.423}&{1.644}
\end{array}} \right)GeV,\\\\
{{\tilde M}_\nu } = \left( {\begin{array}{*{20}{c}}
0&{0.0056 - 0.0111}&0\\
{0.0056 - 0.0111}&{0.0065 - 0.0116}&{0.0223 - 0.0266}\\
0&{0.0223 - 0.0266}&{0.0294 - 0.0390}
\end{array}} \right)eV,
\end{array}
\end{equation}
\end{widetext} 
wherein ${\phi _1} = 0^\circ - 11^\circ ,251^\circ - 360^\circ$, ${\phi _2} = 89^\circ - 268^\circ$ and ${m_{\nu 1}} = (2.26 - 7.53)$ $meV$ , ${m_{\nu 2}} = (8.73 - 11.6)$ $meV$, ${m_{\nu 3}} = (47.7 - 52.0)$ $meV$, $\Sigma  = (59.0 - 71.0)$ $meV$ and $\left\langle {{m_{ee}}} \right\rangle  = (0.01 - 10.0)$ $meV$ respectively. The $m_{\nu 1}$ dependence for $s_{12}^2$ remains the same as before whilst the other two mixing angles being fixed by the parameter $d_\nu$. Furthermore, apart from the phases $\phi_1$ and $\phi_2$, $\delta_l$ is now fixed by the parameter $f_l=\sqrt{-e_l c_l}$ as shown in the FIG. 7. Naturalness in $M'_l$ and $M_\nu$ seems to be in good agreement with $\delta_l\sim270^\circ$ and $s_{23}^2\gtrsim 0.5$ compatible with $ f'_l\sim 0.075$ $GeV$ $ \sim O(\sqrt{m_e m_\mu})< m_\mu$ and $d_\nu \lesssim m_{\nu 2}$ respectively. A greater agreement with naturalness in $M_l$ is achieved for  $\delta_l \longrightarrow 270^\circ \pm 30^\circ$ up to $ f'_l\sim 0.05$ $GeV$.
\subsubsection{Case-C $M_l (11) = M_l (12,21)= 0$}
The viable best-fit structures so obtained for these lepton mass matrices are shown below, 
\begin{widetext}
\begin{equation}\label{D}
\begin{array}{c}
{{\tilde {M}^\prime}_l} = \left( {\begin{array}{*{20}{c}}
0&0&0.029-0.167\\
0&0.003-0.103&{0.395 - 0.580}\\
0.029-0.167&{0.395 - 0.580}&{1.54-1.64}
\end{array}} \right)GeV,\\\\
{{\tilde M}_\nu } = \left( {\begin{array}{*{20}{c}}
0&{0.0022 - 0.0113}&0\\
{0.0022 - 0.0113}&{0.0027 - 0.0117}&{0.0223 - 0.0277}\\
0&{0.0223 - 0.0277}&{0.0283 - 0.0399}
\end{array}} \right)eV,
\end{array}
\end{equation}
\end{widetext} 
wherein ${\phi _1} = 0^\circ - 36^\circ ,175^\circ - 360^\circ$, ${\phi _2} = 97^\circ - 265^\circ$ and ${m_{\nu 1}} = (0.4 - 7.8)$ $meV$ , ${m_{\nu 2}} = (8.4 - 11.7)$ $meV$, ${m_{\nu 3}} = (47.6 - 52.0)$ $meV$, $\Sigma  = (56.9 - 71.2)$ $meV$ and $\left\langle {{m_{ee}}} \right\rangle  = (0.02 - 9.6)$ $meV$  respectively. It is noteworthy that the condition of texture zero at $M'_l (12,21)$ i.e. $a'_l = 0$ fixes the parameter $e_l$ and hence 

\begin{center}
$f'_l=\sqrt{-e_l c_l}$
\end{center}

\begin{flushleft}
through the Eq.(\ref{32}) with 
\end{flushleft}

\begin{center}
$e_l = -m_e m_\mu m_\tau / d_l c_l$.
\end{center}
This results in only one free parameter $d_l = d'_l$ in $M'_l$. This is depicted in FIG.8. This parameter also determines the Dirac-like CP phase as shown in FIG.9. Other observations pertaining to the dependence of mixing angles remain same as previous cases. It is clear that naturalness in $M'_l$ and $M_\nu$ is in good agreement with $\delta_l\sim270^\circ$ and $s_{23}^2\gtrsim 0.5$ compatible with $ f'_l\sim0.064$ $GeV$ $ \sim O(\sqrt{m_e m_\mu})< m_\mu$ and $d_\nu \lesssim m_{\nu 2}$ respectively.
\subsection{Type-II $M_\nu (11) = M_\nu (12,21)= 0$}
\subsubsection{Case-D $M_l (11) = M_\nu (22)= 0$}
The best-fit values obtained for this possibility are summarized below:
\begin{widetext}
\begin{equation}\label{E}
\begin{array}{c}
{{\tilde {M}^\prime}_l} = \left( {\begin{array}{*{20}{c}}
0&0.007-0.094&0.0004-0.220\\
0.007-0.095&0&{0.348 - 0.423}\\
0.0004-0.220&{0.348 - 0.423}&{1.644}
\end{array}} \right)GeV,\\\\
{{\tilde M}_\nu } = \left( {\begin{array}{*{20}{c}}
0&0&0.006-0.019\\
0&{0.0041 - 0.0120}&{0.0178 - 0.0269}\\
0.006-0.019&{0.0178 - 0.0269}&{0.0282 - 0.0393}
\end{array}} \right)eV,
\end{array}
\end{equation}
\end{widetext} 
wherein ${\phi _1} = 0^\circ - 23^\circ ,256^\circ - 360^\circ$, ${\phi _2} = 98^\circ - 261^\circ$ and ${m_{\nu 1}} = (1.1 - 7.9)$ $meV$ , ${m_{\nu 2}} = (8.4 - 12.5)$ $meV$, ${m_{\nu 3}} = (47.6 - 51.9)$ $meV$, $\Sigma  = (57.5 - 70.8)$ $meV$ and $\left\langle {{m_{ee}}} \right\rangle  = (0.01 - 9.56)$ $meV$  respectively. It is observed that naturalness is in good agreement with $\delta_l\sim270^\circ$ and $s_{23}^2\gtrsim 0.5$ compatible with $\mid f'_l\mid\sim0.088$ $GeV$ $ \sim O(\sqrt{m_e m_\mu})< m_\mu$, see FIG.10 and $d_\nu \lesssim m_{\nu 2}$ respectively. Again a greater agreement with naturalness in $M_l$ can be achieved for  $\delta_l \longrightarrow 270^\circ \pm 30^\circ$ up to $\mid f'_l\mid\sim 0.05$ $GeV$. 
\begin{figure}
\includegraphics[scale=1.0]{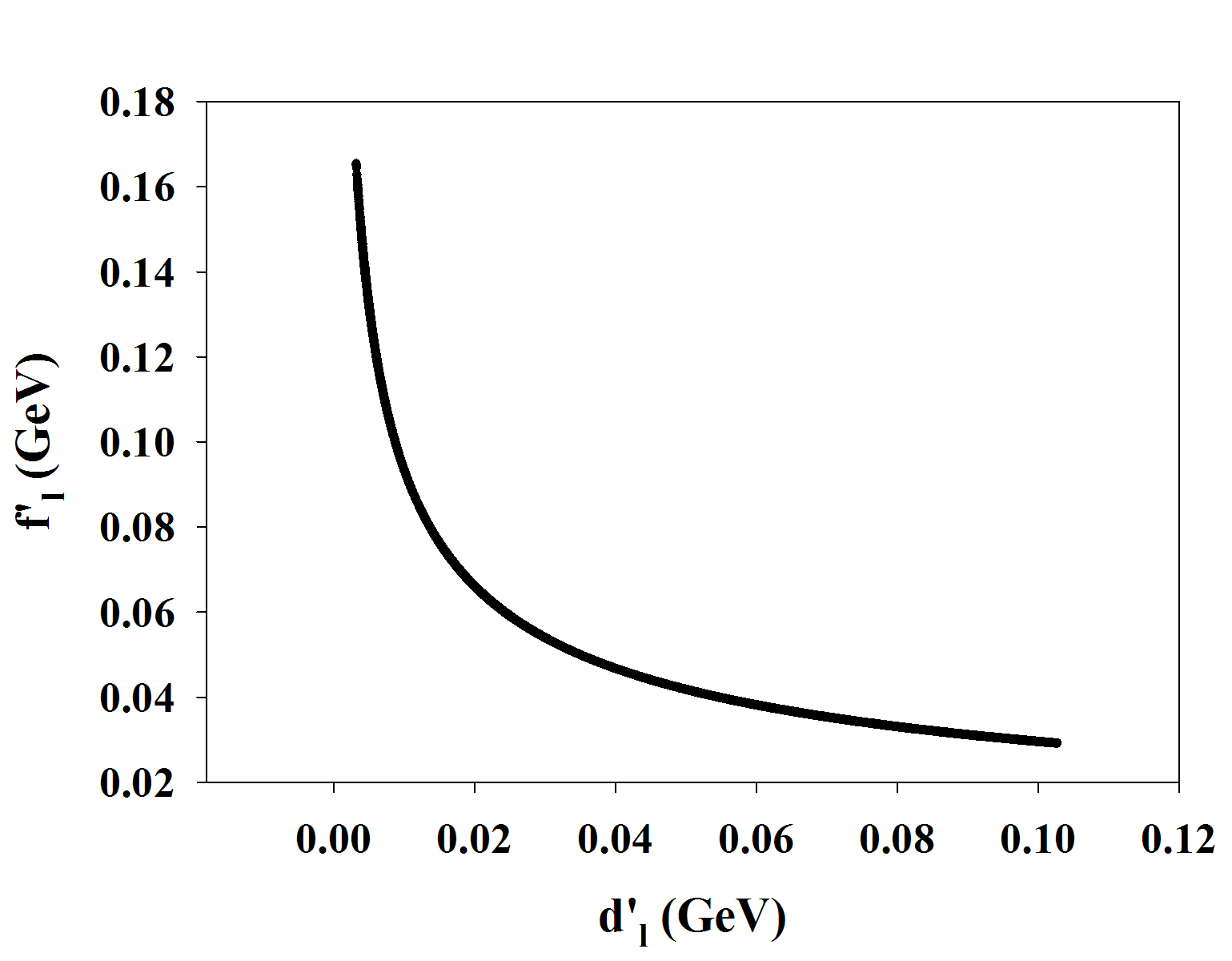}
\caption{$f'_l$ vs. $d'_l$ for Case-C.}
\end{figure}
\begin{figure}
\includegraphics[scale=1.0]{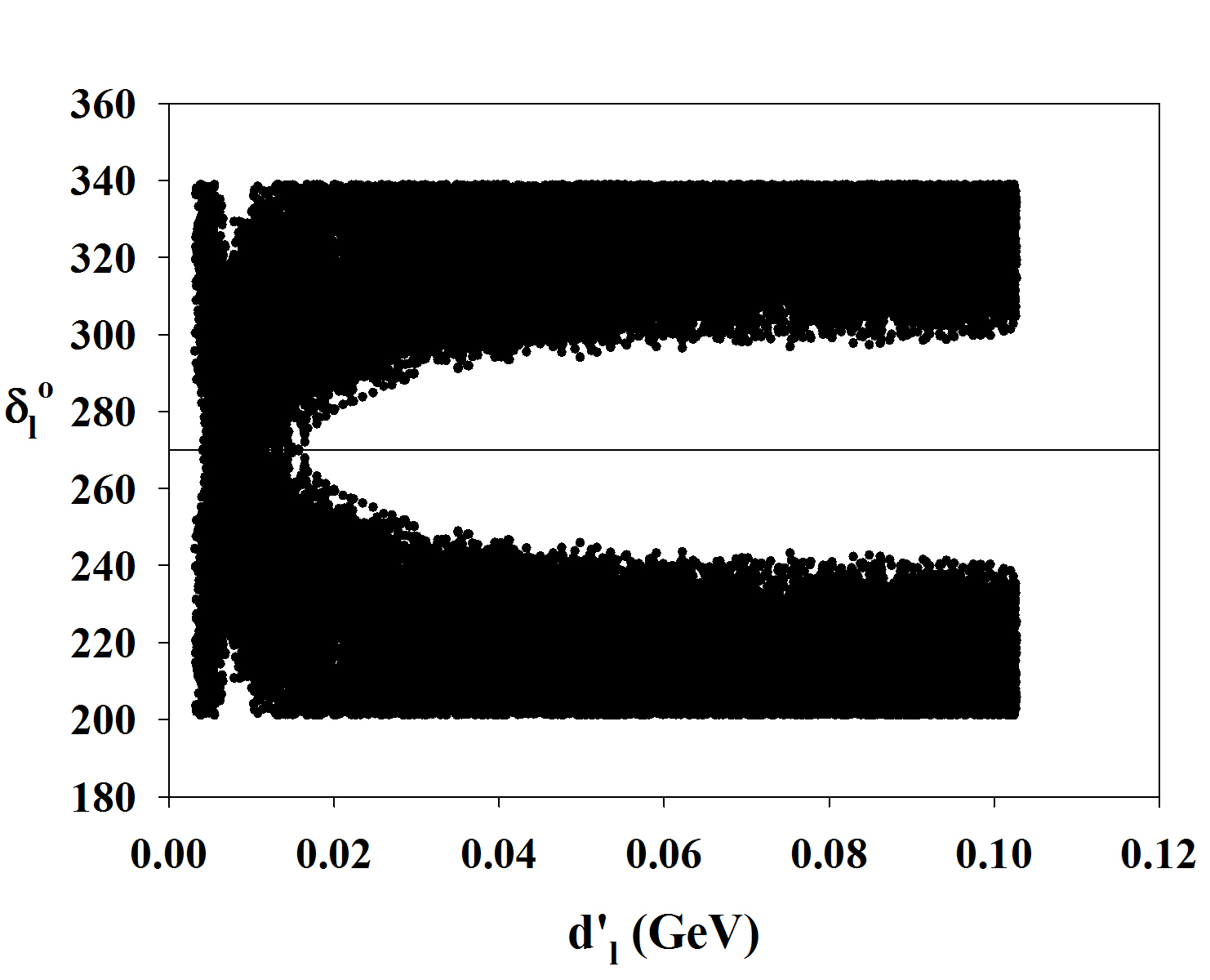}
\caption{$\delta_l$ vs. $d'_l$ for Case-C.}
\end{figure}
\begin{figure}
\includegraphics[scale=1.0]{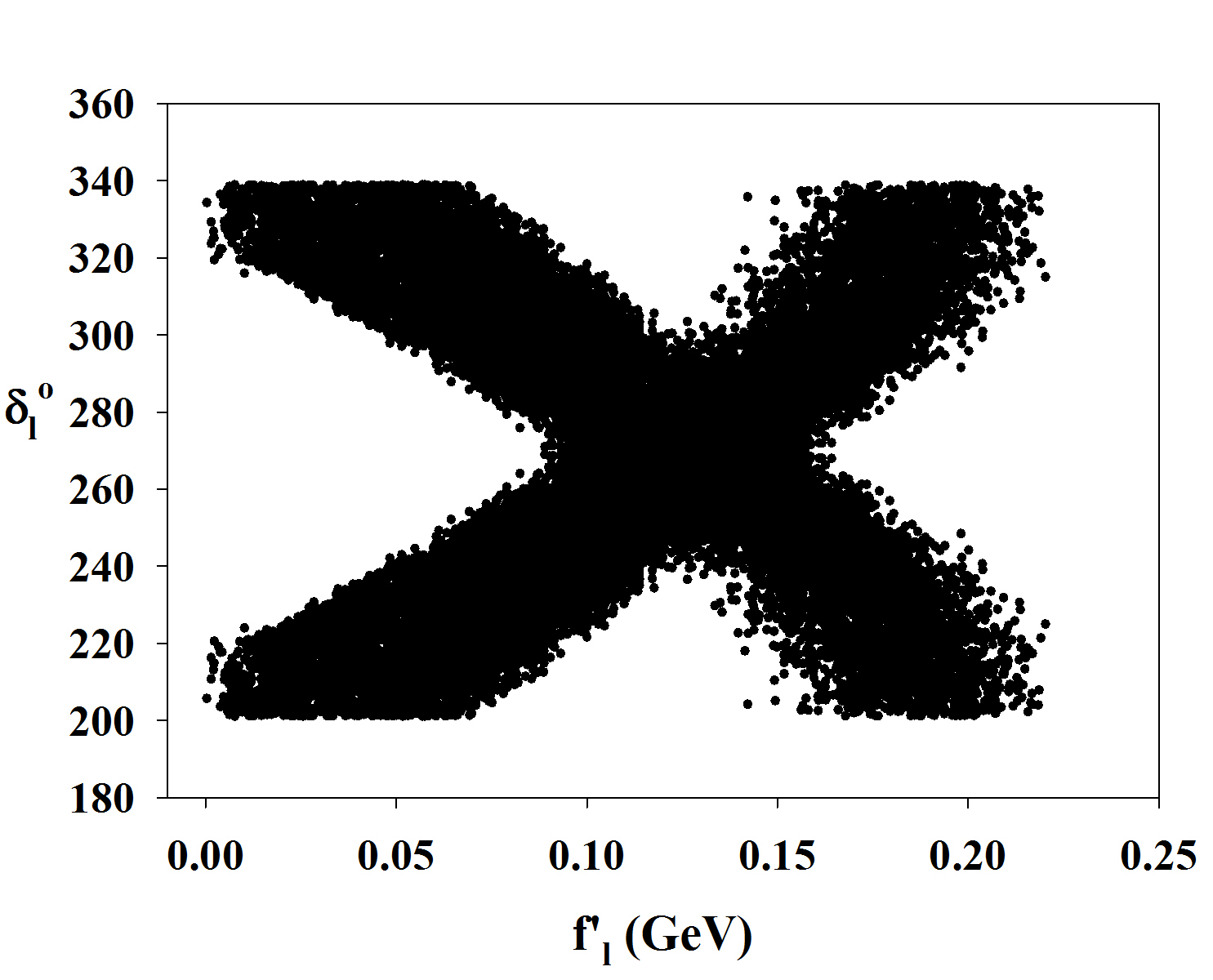}
\caption{$\delta_l$ vs. $f'_l$ for Case-D.}
\end{figure}
\section{Conclusions}
Assuming factorizable phases in lepton mass matrices, we show that natural mass matrices characterized by ${({M_{ij}})} \sim O(\sqrt {{m_i}{m_j}}) $ for $i,j=1,2,3, i\ne j$ and ${({M_{ii}})} \sim O({{m_i}})$ provide a reasonable explanation for the observed fermion masses and flavor mixing patterns in the quark as well as the lepton sectors. It is also observed that deviations from parallel texture structures for $M_{l,d}$ and $M_{\nu,u}$ are essential for establishing such natural structures. Such phenomenological textures have also been observed to be stable under the renormalization group running from the heavy right-handed neutrino mass scale to the electroweak scale \cite{Liao:2015hya, Fukugita2012294, Fukugita2003273, Fukugita01011993, Hagedorn200463}. 

Interestingly, naturalness in the lepton sector implies $s_{12} \propto O(\sqrt{m_{\nu 1}/m_{\nu 2})}$ and $s_{23}^2 \propto d_\nu/c_\nu$ or $s_{23} \propto O(\sqrt{m_{\nu 2}/m_{\nu 3})}$ such that the observed large values of these mixing angles are perhaps indicative of the possible realization of the neutrino mass ratios as obtained above, i.e. ${m_{\nu 1}} \simeq (0.1 - 8.0)$ $meV$, ${m_{\nu 2}} \simeq (8.0 - 13.0)$ $meV$, ${m_{\nu 3}} \simeq (47.0 - 52.0)$ $meV$, $\Sigma  \simeq (56.0 - 71.0)$ $meV$ and $\left\langle {{m_{ee}}} \right\rangle  \simeq (0.01 - 10.0)$ $meV$ respectively. In particular, the possibility of a vanishing neutrino mass i.e. $m_{\nu 1}=0$ is not supported by natural lepton matrices. From the point of view of $0\nu \beta \beta$ decays, these results seem to indicate that multi-ton scale detectors may be required to possibly observe signals for such processes.
\begin{widetext}
\begin{center}
\begin{table}
\begin{tabular}{c c c c c c c c c}
\hline
Sr. & $X_l$ & $X_\nu$ & $ (a) s_{23}^2 \gtrsim 0.5$ & $ (b) \delta_l \sim 270^\circ $ & (c) Natural & (a+c) & (b+c) & Det$M_{\nu}=0$ \\ [0.5ex]
\hline
1 & 11,22,13,31 & 11,13,31 &$\surd$ & $\times$ &$\surd$& $\surd$ & $\times$ & $\times$ \\
\hline
2 & 11,13,31 & 11,22,13,31 &$\surd$ & $\times$ &$\times$& $\times$ & $\times$ & $\times$\\
\hline
3 & 11,13,31,23,32 & 11,13,31 &$\surd$ & $\times$ &$\times$& $\times$ & $\times$ & $\times$ \\
\hline
4 & 12,21,22,13,31 & 11,13,31 &$\surd$ & $\times$ &$\times$& $\times$ & $\times$ & $\times$\\
\hline
5 & 11,13,31,23,32 & 11,12,21 &$\surd$ & $\times$ &$\times$& $\times$ & $\times$ & $\times$\\
\hline
6 & 11,22,13,31 & 11,12,21 &$\surd$ & $\times$ &$\surd$& $\surd$ & $\times$ & $\times$\\
\hline
7 & 12,21,22,13,31 & 11,12,21 &$\surd$ & $\times$ &$\surd$& $\surd$ & $\times$ & $\times$ \\
\hline
8 & 11,12,21,23,32 & 11,13,31 &$\surd$ & $\times$ &$\times$& $\times$ & $\times$  & $\times$\\
\hline
\end{tabular}
\caption{Viable texture five zeros in relation to $s_{23}^2 \gtrsim 0.5$, $ \delta_l \sim 270^\circ $, naturalness and $m_{\nu 1}=0$.}
\end{table}
\begin{table}
\begin{tabular}{c c c c c c c c c}
\hline
Sr. & $X_l$ & $X_\nu$ & $ (a) s_{23}^2 \gtrsim 0.5$ & $ (b) \delta_l \sim 270^\circ $ & (c) Natural & (a+c) & (b+c) & Det$M_{\nu}=0$ \\ [0.5ex]
\hline
1 & 11,13,31 & 11,13,31 &$\surd$ & $\surd$ &$\surd$& $\surd$ & $\times$ & $\times$ \\
\hline
2 & 13,31,23,32 & 11,13,31 &$\surd$ & $\surd$ &$\times$& $\times$ & $\times$ & $\times$\\
\hline
3 & \textbf{11,22} & \textbf{11,13,31} &$\surd$ & $\surd$ &$\surd$& $\surd$ & $\surd$ & $\times$ \\
\hline
4 & 11,13,31 & 11,22 &$\surd$ & $\surd$ &$\surd$& $\times$ & $\times$ & $\surd$\\
\hline
5 & 13,31 & 11,22,13,31 &$\surd$ & $\surd$ &$\surd$& $\times$ & $\surd$ & $\times$\\
\hline
6 & 11,22,13,31 & 13,31 &$\surd$ & $\times$ &$\surd$& $\surd$ & $\times$ & $\surd$\\
\hline
7 & 13,31 & 11,13,31,23,32 &$\surd$ & $\surd$ &$\times$& $\times$ & $\times$ & $\times$ \\
\hline
8 & 11,13,31,23,32 & 13,31 &$\surd$ & $\times$ &$\times$& $\times$ & $\times$ & $\surd$ \\
\hline
9 & 11 & 11,22,13,31 &$\surd$ & $\surd$ &$\surd$& $\times$ & $\surd$  & $\times$\\
\hline
10 & 11,22,13,31 & 11 &$\surd$ & $\times$ &$\surd$& $\surd$ & $\times$ & $\surd$\\
\hline
11 & 11 & 11,13,31,23,32 &$\surd$ & $\surd$ &$\times$& $\times$ & $\times$ & $\times$\\
\hline
12 & 11,13,31,23,32 & 11 &$\surd$ & $\times$ &$\times$& $\times$ & $\times$ & $\times$\\
\hline
13 & \textbf{22,13,31} & \textbf{11,13,31} &$\surd$ & $\surd$ &$\surd$& $\surd$ & $\surd$ & $\times$\\
\hline
14 & \textbf{11,12,21} & \textbf{11,13,31} &$\surd$ & $\surd$ &$\surd$& $\surd$ & $\surd$ & $\times$\\
\hline
15 & 11,13,31 & 11,12,21 &$\surd$ & $\surd$ &$\surd$& $\surd$ & $\times$ & $\times$ \\
\hline
16 & 13,31,23,32 & 11,12,21 &$\surd$ & $\surd$ &$\times$& $\times$ & $\times$ & $\times$ \\
\hline
17 & 22,13,31 & 11,12,21 &$\surd$ & $\surd$ &$\surd$& $\surd$ & $\times$ & $\times$ \\
\hline
18 & 11,12,21 & 11,22 &$\surd$ & $\surd$ &$\times$& $\times$ & $\times$ & $\surd$\\
\hline
19 & \textbf{11,22} & \textbf{11,12,21} &$\surd$ & $\surd$ &$\surd$& $\surd$ & $\surd$ & $\times$\\
\hline
20 & 12,21,13,31 & 11,13,31 &$\surd$ & $\surd$ &$\times$& $\times$ & $\times$ & $\times$\\
\hline
21 & 12,21,13,31 & 11,22 &$\surd$ & $\surd$ &$\times$& $\times$ & $\times$ & $\times$\\
\hline
22 & 22,12,21,13,31 & 13,31 &$\surd$ & $\times$ &$\times$& $\times$ & $\times$ & $\surd$\\
\hline
23 & 12,21,22,13,31 & 11 &$\surd$ & $\times$ &$\surd$& $\surd$ & $\times$  & $\times$\\
\hline
24 & 11,23,32 & 11,13,31 &$\surd$ & $\times$ &$\times$& $\times$ & $\times$ & $\times$ \\
\hline
25 & 11,12,21,23,32 & 11 &$\surd$ & $\times$ &$\times$& $\times$ & $\times$ & $\times$\\
\hline
26 & 11,12,21,23,32 & 13,31 &$\surd$ & $\times$ &$\surd$& $\surd$ & $\times$ & $\times$\\
\hline
27 & 13,31 & 11,12,21,23,32 &$\surd$ & $\surd$ &$\times$& $\times$ & $\times$ & $\times$ \\
\hline
28 & 11 & 11,12,21,23,32 &$\surd$ & $\surd$ &$\times$& $\times$ & $\times$ & $\times$ \\
\hline
\end{tabular}
\caption{Viable texture four zeros in relation to $s_{23}^2 \gtrsim 0.5$, $ \delta_l \sim 270^\circ $, naturalness and Det$M_\nu = 0$.}
\end{table}
\end{center}
\end{widetext}

\begin{acknowledgments}
The author would like to thank Shun Zhou, IHEP, Beijing for discussions and valuable suggestions. This work was supported in part by the Department of Science and Technology under SERB research grant No. SB/FTP/PS-140/2013. 
\end{acknowledgments}
\bibliography{thebibliography}
\end{document}